\documentclass[aps,prb,10pt,twocolumn,superscriptaddress]{revtex4-2}

\usepackage{tikz,pgf}
\usepackage{amsfonts}
\usepackage{amsmath}
\usepackage{amssymb}
\usepackage{amsthm}
\usepackage{bm}
\usepackage{mathtools}
\usepackage{braket}
\usepackage{enumitem}
\usepackage{bbm}
\usepackage{graphicx}
\usepackage{array}
\usepackage{booktabs}
\usepackage{hyperref}
\usepackage{xcolor}
\usepackage{physics}
\usepackage{orcidlink}

\hypersetup{
    colorlinks=true,
    linkcolor=blue,
    citecolor=blue,
    urlcolor=blue
}

\newtheorem{theorem}{Theorem}

\begin{document}

\title{Hybrid Krotov gradient descent Optimization for Quantum Machine Learning}

\author{Maximilian Fröhlich\,\orcidlink{0009-0007-5276-2858}}
\affiliation{Weierstrass Institute for Applied Analysis and Stochastics, Berlin, Germany}
\email{froehlich@wias-berlin.de}
\author{Martin Eigel\,\orcidlink{0000-0003-2687-4497}}
\affiliation{Weierstrass Institute for Applied Analysis and Stochastics, Berlin, Germany}
\author{Michael Hintermüller\,\orcidlink{0000-0001-9471-2479}}
\affiliation{Weierstrass Institute for Applied Analysis and Stochastics, Berlin, Germany}

\date{November 13, 2026}

\begin{abstract}
We introduce and study a Krotov-inspired discrete adjoint optimization scheme for quantum machine learning and focus on a hybrid online-to-batch schedule. The method transfers the forward/costate/local-update logic of continuous-time optimal control to parameterized quantum circuits by replacing time locality with gate-index locality. For a broad class of fixed-generator variational gates and terminal-state losses, we show that the batch Krotov direction coincides with the exact gradient of the empirical loss. Under standard smoothness assumptions, this yields monotonic descent and convergence of the batch method, and convergence of the tail of the hybrid method after a finite switch to the batch phase. Empirically, we benchmark hybrid Krotov against the tuned PennyLane optimizer suite and L-BFGS-B on a four-bit parity classifier and on the Pérez-Salinas data re-uploading classifier for the crown problem. Across matched hyperparameter sweeps with three seeds per configuration, hybrid Krotov dominates every tuned first-order baseline and matches the strongest structured PennyLane methods on parity, while approaching the accuracy of L-BFGS-B on Pérez-Salinas at first-order per-iteration cost.
\end{abstract}

\maketitle

\section{Introduction}

Optimization remains one of the main bottlenecks in variational quantum algorithms and quantum machine learning \cite{CerezoEtAl2021VQA,McCleanEtAl2016Theory,MollEtAl2018QST,BiamonteEtAl2017QML,KandalaEtAl2017HEA,McCleanEtAl2018Barren,WangEtAl2021NoiseBP}. In many settings, standard first-order methods provide robust baselines but may converge slowly or require many objective and gradient evaluations, while second-order and quasi-Newton methods can yield strong late-stage refinement but may be more expensive or less naturally aligned with the structure of parameterized quantum circuits. This raises the question whether optimization rules that explicitly exploit circuit propagation structure can offer a useful alternative.

Krotov's method is a classical optimization technique from optimal control \cite{KonnovKrotov1999,WerschnikGross2007,BrifChakrabartiRabitz2010,PalaoKosloff2003,ReichNdongKoch2012,GoerzEtAl2019Krotov}. Its defining ingredients are a forward state trajectory, a backward sensitivity trajectory, and a local update rule derived from the constrained dynamical structure of the problem. In coherent quantum control, these ingredients match the physics especially naturally: the state evolves unitarily, the costate propagates backward through adjoint dynamics, and the update can be written directly in terms of inner products of propagated states and control generators. This motivates the question whether the same logic can be transferred from continuous-time control to gate-structured quantum machine learning models.

In this work we study a discrete, circuit-level adaptation of this idea and compare it to known optimization procedures used in the realm of quantum machine learning \cite{SchuldEtAl2019Gradients,BanchiCrooks2021SPSR,WierichsEtAl2022ShiftRules}. The central conceptual move is to replace continuous time by gate index. Forward propagation through the circuit produces a sequence of intermediate states, backward propagation produces a sequence of costates, and local gate-wise update directions are constructed from both. We consider three variants: an online version that updates sample by sample, a batch version that averages sample-wise gate contributions over the dataset, and a hybrid method that uses online updates in an initial phase before switching to batch refinement.

Our contributions are both theoretical and empirical. On the theoretical side, we formulate the Krotov-inspired discrete adjoint update as a single object with three clearly distinguished variants and prove that, for fixed-generator variational gates and terminal-state losses, the batch direction coincides with the partial gradient of the empirical loss with respect to the gate-supported parameters. Under standard smoothness assumptions, this yields monotonic descent and convergence of the batch method, and convergence of the tail of the hybrid method after a finite switch to batch updates.
For the online phase, we further identify the implemented update as an inexact sequential coordinate-gradient sweep whose deviation from exact sequential coordinate descent is controlled by the accumulated intra-sample parameter displacement, and we show that under controlled stale-adjoint error each gate update is a descent step for the currently processed sample loss. A direct interpretive consequence, which we develop in detail, is that the fast initial loss decrease characteristic of the hybrid schedule cannot be attributed to a hidden alternative descent direction in the batch phase: the batch direction is the exact full-batch gradient, and the genuinely non-gradient ingredient is the online phase, which performs many local corrections within one epoch through a sequential stale-adjoint sweep.

On the empirical side, we benchmark hybrid Krotov against a broad set of tuned baselines drawn from the \textsc{PennyLane} optimizer suite \cite{BergholmEtAl2018Pennylane}: Adam \cite{KingmaBa2015Adam}, Adagrad \cite{DuchiEtAl2011Adagrad}, RMSProp \cite{TielemanHinton2012RMSProp}, heavy-ball momentum \cite{Polyak1964Momentum}, Nesterov accelerated gradient \cite{Nesterov1983NAG}, plain gradient descent, SPSA \cite{Spall1998SPSA,SpallEtAl2000SPSA}, Rotosolve \cite{OstaszewskiEtAl2021Rotosolve,ParrishEtAl2019JacobiDiag}, and block-diagonal quantum natural gradient \cite{StokesEtAl2020QNG}, together with the SciPy \cite{VirtanenEtAl2020SciPy} implementation of L-BFGS-B \cite{ByrdEtAl1995LBFGSB,ZhuEtAl1997LBFGSB}, on two Quantum machine Learning (QML) classification tasks: the four-bit parity classifier of Farhi and Neven \cite{FarhiNeven2018Classification} and the Pérez-Salinas data re-uploading classifier on the two-dimensional \texttt{crown} problem \cite{PerezSalinasEtAl2020DataReupload}. Across matched hyperparameter sweeps with three seeds per configuration, hybrid Krotov outperforms every tuned \textsc{PennyLane} first-order baseline on both tasks, matches or exceeds the strongest \textsc{PennyLane} contenders on parity, and closes most of the gap to L-BFGS-B on Pérez-Salinas at a fraction of its per-iteration circuit-evaluation cost.

We emphasize from the outset that the present method is not a literal transcription of the classical continuous-time Krotov theorem. The adaptation studied here is best understood as a structured discrete adjoint optimizer for parameterized quantum circuits, and the strongest theoretical guarantees apply to the batch method and to the post-switch phase of the hybrid schedule rather than to the online phase, which is empirically the regime where transient oscillations are observed.

\section{Related Work}

Several lines of prior work bear on the present method. Beyond Krotov's method itself, the quantum optimal control literature has developed a family of continuous-time gradient-based pulse-design algorithms, most prominently gradient ascent pulse engineering (GRAPE) \cite{khaneja2005optimal}, chopped random basis (CRAB) optimization \cite{caneva2011chopped}, and gradient optimization of analytic controls (GOAT) \cite{machnes2018tunable}; comparative studies on gate-preparation tasks find all four to be viable, with the relative ranking depending on the constraint setting \cite{riaz2019optimal}, and recent work has reformulated GRAPE and Krotov within semi-automatic differentiation frameworks that allow arbitrary functionals beyond the standard state-transfer and gate objectives \cite{goerz2022quantum}. All these methods optimize physical pulse amplitudes in continuous time, which is what distinguishes them from the gate-index setting considered here. We note that a sequential-sweep variant of GRAPE on gate index would yield an algorithm closely related to online Krotov, while CRAB and GOAT rely on global pulse parametrizations that have no natural analogue in a finite-dimensional gate-angle space; benchmarking such a sequential-sweep GRAPE variant alongside online Krotov is out of scope for the present work. On the simulation side, the gate-local update formula derived in Sec.~\ref{sec:numerical} is structurally identical to the reverse-mode adjoint scheme of Jones and Gacon \cite{jones2020efficientcalculationgradientsclassical}, which underlies the analytic gradient backends in \textsc{PennyLane} \cite{BergholmEtAl2018Pennylane} and Yao.jl \cite{LuoEtAl2020Yao}; we make this connection explicit in Theorem~\ref{thm:exact_gradient_krotov_main}. A separate strand modifies the descent direction itself by exploiting the geometry of the quantum state manifold: quantum natural gradient (QNG) \cite{StokesEtAl2020QNG} preconditions the gradient with the Fubini-Study metric tensor, with subsequent extensions covering noisy and non-unitary circuits \cite{koczor2022quantum}, an SPSA-style estimator of the quantum Fisher information \cite{GaconEtAl2021SPSAQFI}, generalized variants with adaptive learning rates \cite{haug2104optimal}, and applications to reinforcement-learning policies \cite{meyer2023quantum}. These methods share with hybrid Krotov the aim of exploiting circuit structure rather than treating the loss as a black box, but they use the metric tensor as their structural lever rather than gate-by-gate adjoint propagation. Finally, cross-method optimizer comparisons for variational quantum algorithms have been carried out for chemistry tasks with the unitary coupled-cluster ansatz \cite{singh2023benchmarking}, the Hamiltonian variational ansatz on the Fermi-Hubbard model \cite{jones2025benchmarking}, the variational quantum linear solver \cite{pellow2021comparison}, and a randomized-tasks protocol for local zeroth-order methods \cite{tecot2024randomized}, while classical-side benchmarking has documented the sensitivity of ranking outcomes to hyperparameter tuning protocols \cite{SchmidtEtAl2021DescendingCrowded}; none of these studies includes Krotov-type adjoint sweeps among the optimizer pool, and the QML classification setting is itself comparatively underrepresented in this benchmarking literature.

\section{From continuous-time Krotov to gate-local QML updates}

\subsection{Continuous-time Krotov in optimal control}
\label{sec:continuous_krotov}

Krotov's method originates in constrained optimal control \cite{KonnovKrotov1999} and was subsequently adapted to quantum optimal control, in particular for the design of laser pulses in molecular dynamics \cite{somloi1993controlled, tannor1992control}. In the standard setting, a control $\epsilon:[0,T]\to\mathbb R^m$ drives a state trajectory $\psi:[0,T]\to\mathcal H$ in a Hilbert space $\mathcal H$, and one seeks to minimize a functional of the form
\begin{equation}
\begin{aligned}
J[\psi,\epsilon]
&=
\Phi(\ket{\psi(T)})\\
&\quad +
\int_0^T g_a(\epsilon(t),t)\,dt
+
\int_0^T g_b(\ket{\psi(t)},t)\,dt.
\label{eq:generic_cost}
\end{aligned}
\end{equation}
Here, $\Phi:\mathcal H\to\mathbb R$ is a terminal cost functional that evaluates the final state, while $g_a$ and $g_b$ penalize the control and, optionally, the intermediate state along the trajectory. One typically assumes that $\Phi$ is sufficiently differentiable so that its first variation is well defined, and that $\psi$ is sufficiently regular to satisfy the dynamical constraint.

In coherent quantum dynamics, $\ket{\psi(t)}$ often satisfies
\begin{equation}
i \frac{d}{dt}\ket{\psi(t)}=H(\epsilon(t),t)\ket{\psi(t)},
\label{eq:schrodinger}
\end{equation}
with prescribed initial condition $\ket{\psi(0)}=\ket{\psi_0}$. Krotov's method constructs a new control $\epsilon^{(i+1)}(t)$ from a current iterate $\epsilon^{(i)}(t)$ by combining a forward propagation of the state, a backward propagation of a costate $\ket{\chi(t)} \in \mathcal{H}$, and a local update rule. A typical schematic form is
\begin{equation}
\epsilon^{(i+1)}(t)
=
\epsilon^{(i)}(t)
+
\frac{S(t)}{\lambda_a}
\operatorname{Im}
\left\langle
\chi^{(i)}(t)
\middle|
\frac{\partial H}{\partial \epsilon}
\middle|
\psi^{(i+1)}(t)
\right\rangle,
\label{eq:krotov_update}
\end{equation}
where $S(t)$ is a shape function and $\lambda_a$ is a regularization parameter. The defining conceptual feature is that the update respects the dynamical structure of the problem rather than treating the objective as a black-box function of the control \cite{SklarzTannor2002,PalaoKosloff2003,ReichNdongKoch2012,WerschnikGross2007,BrifChakrabartiRabitz2010}.

\subsection{Parameterized quantum models and terminal-state losses}

In the quantum machine learning setting considered here, the trainable model is a parameterized quantum circuit
\begin{equation}
U(\bm{\theta},x)=U_M(\theta_M,x)\cdots U_1(\theta_1,x),
\label{eq:circuit}
\end{equation}
with trainable parameters $\bm{\theta}=(\theta_1,\dots,\theta_M)$ and input $x$. 
Throughout this work we use a gate-local indexing convention in which each trainable circuit factor carries a single scalar trainable parameter. Thus, in Eq.~\eqref{eq:circuit}, $U_k(\theta_k,x)$ denotes a one-parameter trainable factor. Multi-parameter gates are included after factorization into an ordered product of one-parameter factors.
Additional classical post-processing parameters, when present, are denoted
separately by $\bm{\vartheta}$. Given a fixed initial state $|\psi_0\rangle$, the final state for sample $x_n$ is
\begin{equation}
|\psi_{n,M}(\bm{\theta})\rangle
=
U(\bm{\theta},x_n)|\psi_0\rangle.
\label{eq:final_state}
\end{equation}

We consider supervised binary classification with an empirical terminal-state
loss
\begin{equation}
\begin{aligned}
F(\bm{\theta},\bm{\vartheta})
&=
\frac{1}{N}\sum_{n=1}^N L_n(\bm{\theta},\bm{\vartheta}),
\\
L_n(\bm{\theta},\bm{\vartheta})
&=
\ell(p_n(\bm{\theta},\bm{\vartheta}),y_n),
\label{eq:empirical_loss}
\end{aligned}
\end{equation}
where $p_n$ denotes the scalar model readout for sample $n$, not necessarily a probability. If no classical post-processing parameters are
present, we omit $\bm{\vartheta}$ from the notation.
Depending on the benchmark, labels may be encoded either as
$y_n\in\{0,1\}$, together with a probabilistic readout and a loss such as
binary cross-entropy, or as $y_n\in\{-1,+1\}$, together with an
expectation-value readout and a loss such as mean squared error. The abstract
theory below only requires that each $L_n$ be differentiable as a function of
the final circuit state \cite{MitaraiEtAl2018QCL,SchuldEtAl2020CircuitCentric,PerezSalinasEtAl2020DataReupload}.

One common probabilistic one-observable readout is obtained from
\begin{equation}
z_n(\bm{\theta})
=
\langle \psi_{n,M}(\bm{\theta})|O|\psi_{n,M}(\bm{\theta})\rangle,
\qquad
p_n(\bm{\theta})=\frac{z_n(\bm{\theta})+1}{2}.
\label{eq:simple_readout}
\end{equation}
If $\|O\|_{\mathrm{op}}\leq 1$, then $p_n\in[0,1]$, so this readout is
compatible with $y_n\in\{0,1\}$ losses such as binary cross-entropy.
Here $\| \cdot \|_{\mathrm{op}}$ states the operator norm. For expectation-value based losses, as in the parity benchmark below, one may
instead use the scalar prediction $p_n=z_n$ directly with labels
$y_n\in\{-1,+1\}$.

More generally, if the circuit output is passed into a classical affine head,
we write $\bm{\vartheta}=(\bm w,b)$ and set
\begin{align}
&m_{n,j}(\bm{\theta})
=
\langle \psi_{n,M}(\bm{\theta})|O_j|\psi_{n,M}(\bm{\theta})\rangle,\\
&s_n(\bm{\theta},\bm{\vartheta})
=
\bm{w}^{\top}\bm{m}_n(\bm{\theta})+b=\sum_{j=1}^J w_j m_{n,j}(\bm{\theta})+b,
\\
&p_n(\bm{\theta},\bm{\vartheta})=\sigma(s_n).
\label{eq:hybrid_head}
\end{align}

Here $O_j$ denotes the j-th observable.
The loss derivative then propagates through the classical head before defining the terminal costate \cite{SchuldEtAl2019Gradients,BanchiCrooks2021SPSR,WierichsEtAl2022ShiftRules}. Here $\sigma$ is a classical activation function mapping $s_n$ onto the relevant interval for the classification problem.

\subsection{Gate-local forward and backward propagation}
\label{sec:gate-local-forward-backward}

Motivated by the forward-state/costate structure of Krotov methods in
continuous-time quantum optimal control
\cite{KonnovKrotov1999, GoerzEtAl2019Krotov, ReichNdongKoch2012},
we obtain a discrete analogue by replacing time locality with gate-index
locality.

For each sample $n$, define forward states
\begin{equation}
\begin{aligned}
|\psi_{n,0}\rangle&=|\psi_0\rangle,\\
|\psi_{n,k}(\bm{\theta})\rangle
&=
U_k(\theta_k,x_n)\cdots U_1(\theta_1,x_n)|\psi_0\rangle,
\label{eq:forward_states}
\end{aligned}
\end{equation}
for $k=1,\dots,M$.

First consider the case without a classical post-processing head. In analogy to the continuous-time setting in \ref{sec:continuous_krotov}, we also define a terminal costate. The role of
the terminal costate is to represent the derivative of the sample loss with
respect to the final quantum state.

Although $L_n$ was introduced as a function of the parameters
$(\bm{\theta},\bm{\vartheta})$, its dependence on the quantum circuit enters
only through the final state $\ket{\psi_{n,M}(\bm{\theta})}$ in the case without classical post-processing head. In the
following we use the
same symbol $L_n$ also for the corresponding terminal state functional.

For an infinitesimal perturbation of the final state,
\[
|\psi_{n,M}(\bm{\theta})\rangle
\mapsto
|\psi_{n,M}(\bm{\theta})\rangle
+
|\delta\psi_{n,M}(\bm{\theta})\rangle,
\]
the first variation $\delta L_n$ is the linear term in the expansion

\begin{equation}
\begin{aligned}
&L_n(\ket{\psi_{n,M}(\bm{\theta})}+\ket{\delta\psi_{n,M}(\bm{\theta})})\\
&=
L_n(\ket{\psi_{n,M}(\bm{\theta})})+
\delta L_n
+
o(\|\ket{\delta\psi_{n,M}(\bm{\theta})}\|_{\mathcal{H}}).
\end{aligned}
\end{equation}
We define the terminal costate
$|\chi_{n,M}(\bm{\theta})\rangle \in \mathcal{H}$ as the vector representing this
linear functional, i.e. 
\begin{equation}
\delta L_n(\bm{\theta})
=2\mathrm{Re}
\langle
\chi_{n,M}(\bm{\theta})
\mid
\delta\psi_{n,M}(\bm{\theta})
\rangle .
\label{eq:first_variation}
\end{equation}
for arbitrary first-order variations
$|\delta\psi_{n,M}(\bm{\theta})\rangle$ of the final state. 
Thus $|\chi_{n,M}\rangle$ is the terminal adjoint variable: it encodes how an
infinitesimal perturbation of the final circuit state changes the sample loss.

For the simple one-observable readout
\begin{equation}
p_n(\bm{\theta})=\frac{
\langle
\psi_{n,M}(\bm{\theta})
|
O
|
\psi_{n,M}(\bm{\theta})
\rangle
+1
}{2},
\end{equation}
with $O=O^\dagger$, the chain rule gives
\begin{align}
\delta L_n
&=
\frac{\partial \ell}{\partial p_n}\delta p_n,\
\qquad\delta p_n
=\mathrm{Re}
\langle
O\psi_{n,M}(\bm{\theta})
\mid
\delta\psi_{n,M}(\bm{\theta})
\rangle .
\end{align}
Comparing this expression with the defining identity
\eqref{eq:first_variation}, we obtain
\begin{equation}
|\chi_{n,M}(\bm{\theta})\rangle=\frac{1}{2}
\frac{\partial \ell}{\partial p_n}
O|\psi_{n,M}(\bm{\theta})\rangle .
\label{eq:terminal_costate_simple}
\end{equation}
The prefactor $1/2$ is a consequence of the normalization
of $p_n$. Other terminal readouts are treated in the
same way by applying the chain rule to the final-state dependence of $L_n$.
For an affine classical head, the head parameters are denoted by
$\bm{\vartheta}=(\bm w,b)$. In the gate-local Krotov construction these
parameters are treated as fixed when differentiating with respect to the
quantum gate parameters $\bm{\theta}$. No factor $1/2$ is absorbed into the
affine readout. Setting
\begin{equation}
A(\bm{\vartheta})
:=
\sum_j w_j O_j ,
\end{equation}

we have, for fixed $\bm{\vartheta}$,
\begin{align}
\delta s_n
&=
\sum_j w_j\delta m_{n,j} \
&=
2\mathrm{Re}
\langle
A(\bm{\vartheta})\psi_{n,M}
\mid
\delta\psi_{n,M}
\rangle .
\end{align}

By the scalar chain rule applied to
$L_n=\ell(\sigma(s_n),y_n)$, define the loss sensitivity with respect to the
affine score by
\[
\beta_n(\bm{\theta},\bm{\vartheta})
:=
\frac{\partial L_n}{\partial s_n}
=\frac{\partial \ell}{\partial p_n}
\frac{\partial p_n}{\partial s_n}
=\frac{\partial \ell}{\partial p_n}\sigma'(s_n).
\]
Thus
\[
\delta L_n
=
\beta_n(\bm{\theta},\bm{\vartheta})\delta s_n.
\]

Inserting $\delta s_n$ then yields
\[
\delta L_n
=2\mathrm{Re}
\langle
\beta_n(\bm{\theta},\bm{\vartheta})
A(\bm{\vartheta})
\psi_{n,M}
\mid
\delta\psi_{n,M}
\rangle .
\]
Comparing again with \eqref{eq:first_variation} gives the terminal costate
\begin{equation}
|\chi_{n,M}(\bm{\theta},\bm{\vartheta})\rangle
=\beta_n(\bm{\theta},\bm{\vartheta})
A(\bm{\vartheta})
|\psi_{n,M}(\bm{\theta})\rangle .
\label{eq:terminal_costate_affine}
\end{equation}
In the common case of a sigmoid output with binary cross-entropy loss,
$\beta_n=p_n-y_n$, and therefore
\begin{equation}
|\chi_{n,M}(\bm{\theta},\bm{\vartheta})\rangle
=(p_n-y_n)
\left(
\sum_j w_j O_j
\right)
|\psi_{n,M}(\bm{\theta})\rangle .
\label{eq:terminal_costate_affine_bce}
\end{equation}
When no classical head is present, the dependence on
$\bm{\vartheta}$ is omitted.
Backward-propagated costates are then defined by

\begin{equation}
\begin{aligned}
&|\chi_{n,k}^{\mathrm{after}}(\bm{\theta},\bm{\vartheta})\rangle
\\&=
U_{k+1}(\theta_{k+1},x_n)^\dagger \cdots U_M(\theta_M,x_n)^\dagger
|\chi_{n,M}(\bm{\theta},\bm{\vartheta})\rangle.
\label{eq:backward_costates}
\end{aligned}
\end{equation}

This construction parallels adjoint propagation in quantum optimal control
\cite{ReichNdongKoch2012,GoerzEtAl2019Krotov} and the gate-local gradient identity is closely related to the reverse-mode
state-vector differentiation scheme of Jones and Gacon
\cite{jones2020efficientcalculationgradientsclassical}, which treats
expectation-value objectives. Here we use the same adjoint propagation
principle, but initialize the terminal adjoint with the first variation of the
sample-wise terminal loss $L_n=\ell(p_n(\bm{\theta},\bm{\vartheta}),y_n)$. This incorporates the readout map, the loss derivative, and, when present,
the fixed classical head parameters $\bm{\vartheta}$ before the backward
circuit sweep.

\subsection{Online, batch, and hybrid Krotov}

We assume that, after a suitable factorization of the circuit, every trainable gate is parameterized by a single scalar $\theta_k$ and takes the form
\begin{equation}
U_k(\theta_k, x_n) = \exp\!\Bigl(-\tfrac{i}{2}\,\alpha_k(\theta_k, x_n)\,G_k\Bigr),
\label{eq:fixed_generator_gate}
\end{equation}

where $G_k$ is a fixed Hermitian generator independent of $\theta_k$, and
$\alpha_k:\mathbb R\times\mathcal X\to\mathbb R$ is a scalar angle map differentiable with respect to $\theta_k$. Multi-parameter gates such as Euler rotations $R_z(\alpha)R_y(\beta)R_z(\gamma)$ are admissible under this assumption only after being written as an ordered product of single-generator factors, which is the convention adopted throughout this paper and in all benchmarks of Sec.~\ref{sec:numerical}.

The corresponding gate-local contribution for sample $n$ is
\begin{equation}
\begin{aligned}
\Delta_{n,k}(\bm{\theta})
&=
\frac{\partial \alpha_k}{\partial \theta_k}(\theta_k,x_n)\\
&\quad {}\times 2\,\mathrm{Re}\!\left\langle
\chi_{n,k}^{\mathrm{after}}(\bm{\theta})
\middle|
\left(-\frac{i}{2}G_k\right)
\middle|
\psi_{n,k}(\bm{\theta})
\right\rangle.
\label{eq:discrete_contribution}
\end{aligned}
\end{equation}

This can be interpreted as a generalization of the reverse-mode gradient derived in \cite{jones2020efficientcalculationgradientsclassical} and yields the same computational advantage in full quantum state simulation compared to the parameter-shift rule.

For fixed $\bm{\theta}$, the quantity $\Delta_{n,k}(\bm{\theta})$ is exactly the sample-wise partial derivative $\partial L_n/\partial \theta_k$. The distinction between the online and batch variants is therefore not the local formula itself, but how this formula is reused during an update sweep.

We consider three algorithmic variants.

\paragraph{Online Krotov.}

For one training sample $(x_n,y_n)$, let
\[
\bm{\theta}^{[0]} := \bm{\theta}^{(m)}
\]
denote the pre-sweep parameter vector of the $m$-th iteration. At this parameter vector we compute the
terminal costate and the backward-propagated costates
\[
|\chi_{n,k}^{\mathrm{after}}(\bm{\theta}^{[0]})\rangle,
\qquad k=1,\dots,M .
\]
These costates are kept fixed during the sample sweep.

The parameters are then updated gate by gate. Immediately before the gate-$k$
update, the current partially updated parameter vector is
$\bm{\theta}^{[k-1]}$. Thus $\theta_1,\dots,\theta_{k-1}$ have already
been modified by the same sample, whereas $\theta_k,\dots,\theta_M$ still
coincide with their pre-sweep values. The online signal is defined by
\begin{equation}
\begin{aligned}
&\widehat\Delta_{n,k}\\
&=
\frac{\partial \alpha_k}{\partial \theta_k}(\theta_k^{[k-1]},x_n)\\
&\quad {}\times 2\,\mathrm{Re}
\left\langle
\chi_{n,k}^{\mathrm{after}}(\bm{\theta}^{[0]})
\middle|
\left(-\frac{i}{2}G_k\right)
\middle|
\psi_{n,k}(\bm{\theta}^{[k-1]})
\right\rangle .
\label{eq:online_mixed_signal}
\end{aligned}
\end{equation}
The corresponding coordinate update is
\begin{equation}
\bm{\theta}^{[k]}
=
\bm{\theta}^{[k-1]}
-
\eta_{\mathrm{on}}\widehat\Delta_{n,k}\bm e_k,
\qquad k=1,\dots,M ,
\label{eq:online_partial_vectors}
\end{equation}
with $\eta_{\mathrm{on}}$ as the step size of the online phase. After one sample sweep of the $m$-th iteration, the updated parameter vector is
\[
\bm{\theta}^{(m+1)}:=\bm{\theta}^{[M]}.
\]
Thus the backward trajectory is fully stale, whereas the forward state
$|\psi_{n,k}(\bm{\theta}^{[k-1]})\rangle$ uses the already updated 
parameters on gates $1, \dots, k-1$ and the pre-update value of the current gate $k$. This variant is
aggressive and often effective in the early training regime, but it may induce
substantial late-stage oscillations.

\paragraph{Batch Krotov.}
We average the sample-wise contributions over the dataset,
\begin{equation}
\bar{\Delta}_k(\bm{\theta})
=
\frac{1}{N}\sum_{n=1}^N \Delta_{n,k}(\bm{\theta}),
\label{eq:batch_direction}
\end{equation}
and perform a single parameter update
\begin{equation}
\theta_k^{(m+1)}=\theta_k^{(m)}-\eta\,\bar{\Delta}_k(\bm{\theta}^{(m)}).
\label{eq:batch_update}
\end{equation}
This is more stable than the online rule and aligns directly with the empirical loss.

We emphasize that the batch variant defined above reduces, by Theorem~\ref{thm:exact_gradient_krotov_main}, to standard full-batch gradient descent on the gate-supported parameters; its Krotov flavor lies in the gate-local construction of the per-sample contributions $\Delta_{n,k}$, not in the aggregated update direction itself. The genuinely non-gradient ingredient of the method is the online phase, whose stale-adjoint structure is analyzed separately in Sec.~\ref{sec:online_explanation}.

\paragraph{Hybrid Krotov.}
The hybrid method uses online Krotov for an initial phase $m<T_{\mathrm{sw}}$ and switches to batch Krotov for $m\geq T_{\mathrm{sw}}$. The intended effect is to combine rapid basin entry with stable late-stage refinement.\\

Two further implementation remarks apply across all three variants. First, for hybrid quantum-classical models, purely classical parameters such as $\bm w$ and $b$ in \eqref{eq:hybrid_head} are not updated by the gate-local Krotov rule; in the experiments below, such parameters receive their exact loss-gradient update within the same training epoch. Second, per-gate scaling factors $S_k$ analogous to $S(t)/\lambda_a$ in the continuous-time update were tested but did not consistently improve performance and are not used in the benchmarks reported below.

\section{Theoretical properties}

\subsection{Monotonicity and convergence of batch krotov}

We now make precise what can be guaranteed for the batch part of the
gate-local Krotov construction. The statements in this subsection are written
for the case without trainable classical post-processing parameters, so that
the empirical loss is denoted by $F(\bm{\theta})$. If a classical affine head
is present but kept fixed during the quantum-parameter update, the same
gate-local argument applies with the corresponding terminal costate from
Sec.~\ref{sec:gate-local-forward-backward}. Purely classical head parameters,
when present, are updated separately by their ordinary loss gradients.

The first result shows that, under the fixed-generator assumption, the
gate-local batch direction is not merely Krotov-inspired but coincides exactly
with the full-batch gradient with respect to the trainable gate parameters.
The second result then records the standard descent and convergence
consequences of this identification under the usual smoothness assumptions.
These guarantees apply to the batch method and, after a finite switch, to the
batch tail of the hybrid method. They do not assert monotonicity for the
online stale-adjoint phase, which is analyzed separately below.

\begin{theorem}[Exact gradient representation of the batch Krotov direction]
\label{thm:exact_gradient_krotov_main}
Let $F(\bm{\theta})$ be defined by \eqref{eq:empirical_loss} and assume that each trainable gate has the fixed-generator form \eqref{eq:fixed_generator_gate} and assume that each sample loss depends on the circuit only through the final state. Then the batch Krotov direction $\bar{\Delta}(\bm{\theta})$ defined by \eqref{eq:batch_direction} coincides with the partial
gradient of the empirical loss with respect to the gate-supported parameters:
\[
\bar\Delta_k(\bm{\theta}) = \frac{\partial F}{\partial \theta_k}(\bm\theta), \qquad k=1,\dots,M.
\]
\end{theorem}

The next result is therefore not intended as a new convergence theorem for gradient descent. Rather, once Theorem~\ref{thm:exact_gradient_krotov_main} identifies the batch Krotov direction with the exact empirical-loss gradient, the standard smooth-gradient descent theory applies directly. We state the result explicitly in order to make clear which monotonicity and convergence guarantees are inherited by the batch Krotov method, and hence by the post-switch tail of the hybrid method. 

\begin{theorem}[Monotonic descent and convergence of batch Krotov]
\label{thm:batch_krotov_convergence_main}
Assume that the batch Krotov direction equals $\nabla F$,  that $F$ is bounded from below and that its gradient is Lipschitz
continuous on the sublevel set
\[
\mathcal S_0
=
\{\bm{\theta}:F(\bm{\theta})\leq F(\bm{\theta}^{(0)})\},
\]
i.e. there exists $L>0$ such that
\[
\|\nabla F(\bm{\theta})-\nabla F(\bm{\zeta})\|
\leq
L\|\bm{\theta}-\bm{\zeta}\|
\qquad
\forall\,\bm{\theta},\bm{\zeta}\in\mathcal S_0.
\]

Consider the batch iteration
\[
\bm{\theta}^{(m+1)}=\bm{\theta}^{(m)}-\eta \nabla F(\bm{\theta}^{(m)})
\]
with fixed step size $0<\eta<2/L$. Then $F(\bm{\theta}^{(m)})$ is monotonically nonincreasing and
\[
F(\bm{\theta}^{(m+1)})
\leq
F(\bm{\theta}^{(m)})
-\eta\left(1-\frac{L\eta}{2}\right)\|\nabla F(\bm{\theta}^{(m)})\|^2.
\]
Moreover,
\[
\sum_{m=0}^{\infty}\|\nabla F(\bm{\theta}^{(m)})\|^2<\infty,
\qquad
\|\nabla F(\bm{\theta}^{(m)})\|\to 0,
\]
and every accumulation point is stationary. If, in addition, the iterates are bounded and $F$ satisfies the Kurdyka-\L{}ojasiewicz property, then the whole sequence converges to a single stationary point.
\end{theorem}
The same conclusion applies to the tail of a hybrid run once the method has
switched from online updates to the batch rule after a finite number of
iterations. Indeed, after the switch the sequence is simply a batch Krotov
sequence initialized at the switching point, so Theorem~\ref{thm:batch_krotov_convergence_main}
applies from that point onward.

The proofs are given in Appendix~A and Appendix~B. The practical interpretation is important. The theoretical guarantees established here apply to the batch method and to the post-switch phase of the hybrid method. They do \emph{not} imply a monotonic-descent theorem for the online stale-adjoint phase, which empirically is precisely the regime where oscillations are observed.

\subsection{Online Krotov as an inexact sequential descent method}
\label{sec:online_explanation}

Theorem~\ref{thm:exact_gradient_krotov_main} shows that the batch Krotov direction is the exact full-batch gradient on the gate-supported parameters. Hence the early empirical advantage of the hybrid method cannot be attributed to a hidden alternative full-batch descent direction. The remaining question is more specific: what is the mathematical structure of the online sweep defined in Eq.~\eqref{eq:online_mixed_signal}?

In this section we show that the online phase is an inexact sequential coordinate-gradient method. The inexactness is caused precisely by the reuse of backward costates computed at the pre-sweep parameter vector, while the forward state is refreshed gate by gate. Under a local smoothness assumption, the resulting stale-adjoint error is controlled by the accumulated intra-sample parameter displacement. Consequently, for sufficiently small online step size and controlled stale-adjoint error, each gate update is a descent step for the currently processed sample loss. This is a sample-wise statement and should not be read as a monotonicity theorem for the full empirical loss.

Fix a training sample $n$, and use the partially updated vectors
$\bm{\theta}^{[k]}$ defined in Eq.~\eqref{eq:online_partial_vectors}.
Immediately before the gate-$k$ update, the parameters
$\theta_1,\dots,\theta_{k-1}$ have already been modified by the same sample,
whereas $\theta_k,\dots,\theta_M$ still coincide with their pre-sweep values.

The exact sequential coordinate-gradient signal at this partially updated point would be
\begin{equation}
g_{n,k}^{\mathrm{seq}}
:=
\frac{\partial L_n}{\partial \theta_k}(\bm{\theta}^{[k-1]}).
\label{eq:exact_sequential_coordinate_gradient}
\end{equation}
The implemented online Krotov signal $\widehat{\Delta}_{n,k}$, however, uses the forward state associated with $\bm{\theta}^{[k-1]}$ up to gate $k$, but the backward costate computed at $\bm{\theta}^{[0]}$. To make this distinction explicit, define the mixed gate-local adjoint signal
\begin{equation}
\begin{aligned}
\mathcal{D}_{n,k}(\bm{a},\bm{b})
&:=
\frac{\partial \alpha_k}{\partial \theta_k}(b_k,x_n)\\
&\quad {}\times 2\,\mathrm{Re}
\left\langle
\chi_{n,k}^{\mathrm{after}}(\bm{a})
\middle|
\left(-\frac{i}{2}G_k\right)
\middle|
\psi_{n,k}(\bm{b})
\right\rangle .
\label{eq:mixed_adjoint_signal}
\end{aligned}
\end{equation}
Here $\bm a$ specifies the parameter vector from which the terminal costate and backward-propagated costate are computed, while $\bm b$ specifies the parameter vector
used for the forward state through gate $k$. By the exact-gradient identity of Theorem~\ref{thm:exact_gradient_krotov_main}, the diagonal value satisfies
\begin{equation}
\mathcal{D}_{n,k}(\bm{b},\bm{b})
=
\frac{\partial L_n}{\partial \theta_k}(\bm{b}).
\label{eq:mixed_signal_diagonal_identity}
\end{equation}
The online Krotov signal can therefore be written as
\begin{equation}
\widehat{\Delta}_{n,k}
=
\mathcal{D}_{n,k}(\bm{\theta}^{[0]},\bm{\theta}^{[k-1]}),
\label{eq:online_signal_as_mixed_signal}
\end{equation}
whereas the exact sequential coordinate gradient is
\begin{equation}
g_{n,k}^{\mathrm{seq}}
=
\mathcal{D}_{n,k}(\bm{\theta}^{[k-1]},\bm{\theta}^{[k-1]}).
\label{eq:exact_seq_gradient_as_mixed_signal}
\end{equation}

The difference between these two quantities is therefore entirely due to the
stale choice of the costate-generating parameter vector. The next result makes
this statement quantitative. It shows that, under a local Lipschitz regularity
condition on the mixed adjoint signal with respect to its first argument, the
online Krotov signal is the exact sequential coordinate-gradient signal plus a
scalar stale-adjoint error controlled by the accumulated intra-sample
parameter displacement.

\begin{theorem}[Stale-adjoint representation of the online Krotov signal]
\label{thm:online_stale_adjoint_representation}
Fix a training sample $n$ and a gate index $k$, and let
$\bm{\theta}^{[0]},\dots,\bm{\theta}^{[k-1]}$ be the partially updated
parameter vectors generated by the online sweep. Suppose that, along this
sweep, the mixed adjoint signal $\mathcal{D}_{n,k}$ is locally Lipschitz in
the parameter vector from which the costate is computed. More precisely,
assume that there exists a constant $C_{n,k}>0$ such that
\begin{equation}
\left|
\mathcal{D}_{n,k}(\bm{a},\bm{b})
-
\mathcal{D}_{n,k}(\bm{b},\bm{b})
\right|
\leq
C_{n,k}\|\bm{a}-\bm{b}\|
\label{eq:mixed_adjoint_lipschitz}
\end{equation}
for all parameter vectors $\bm a,\bm b$ in a neighbourhood containing the
online sweep trajectory. Then the online Krotov signal satisfies
\begin{equation}
\widehat{\Delta}_{n,k}
=
\frac{\partial L_n}{\partial \theta_k}(\bm{\theta}^{[k-1]})
+
e_{n,k},
\qquad
e_{n,k}\in\mathbb R,
\label{eq:online_signal_exact_plus_error}
\end{equation}
where the stale-adjoint error obeys
\begin{equation}
|e_{n,k}|
\leq
C_{n,k}
\|\bm{\theta}^{[k-1]}-\bm{\theta}^{[0]}\|.
\label{eq:stale_error_bound_displacement}
\end{equation}
In particular, for the unscaled online update
\eqref{eq:online_partial_vectors},
\begin{equation}
|e_{n,k}|
\leq
C_{n,k}\eta_{\mathrm{on}}
\left(
\sum_{j=1}^{k-1}
|\widehat{\Delta}_{n,j}|^2
\right)^{1/2}.
\label{eq:stale_error_bound_stepsize}
\end{equation}
\end{theorem}

The estimate has two immediate consequences. For $k=1$, no previous gate has
been updated within the sample sweep, so
$\bm{\theta}^{[k-1]}=\bm{\theta}^{[0]}$ and the stale-adjoint error vanishes.
For later gates, the error is controlled by the distance travelled by the
earlier intra-sample updates. Hence, if the gate-local signals remain bounded,
the online rule approaches an exact sequential coordinate-gradient sweep in the
small-step limit $\eta_{\mathrm{on}}\to 0$.

We now turn this perturbation estimate into a descent statement for the
processed sample. The next theorem combines coordinate-wise smoothness with a
relative stale-adjoint error condition and gives a sufficient condition under
which the full online sweep decreases the current sample loss.

\begin{theorem}[Sample-wise descent of the online Krotov sweep]
\label{thm:online_samplewise_descent}
Fix a training sample $n$ and consider one online sweep over this sample.
Suppose that, for each gate $k=1,\dots,M$, the sample loss $L_n$ is
coordinate-wise $L_{n,k}$-smooth in $\theta_k$ along the sweep, i.e.
\begin{equation}
L_n(\bm{\theta}+s\bm{e}_k)
\leq
L_n(\bm{\theta})
+
s\frac{\partial L_n}{\partial \theta_k}(\bm{\theta})
+
\frac{L_{n,k}}{2}s^2
\label{eq:coordinate_smoothness}
\end{equation}
for all relevant $\bm{\theta}$ and all $s\in\mathbb R$. For each gate, write
\[
g_{n,k}^{\mathrm{seq}}
=
\frac{\partial L_n}{\partial \theta_k}(\bm{\theta}^{[k-1]}),
\qquad
\widehat{\Delta}_{n,k}
=
g_{n,k}^{\mathrm{seq}}+e_{n,k},
\]
where $e_{n,k}\in\mathbb R$ denotes the stale-adjoint error. Assume that this
error satisfies the relative bound
\begin{equation}
|e_{n,k}|
\leq
\rho_{n,k}|g_{n,k}^{\mathrm{seq}}|,
\qquad
0\leq \rho_{n,k}<1.
\label{eq:relative_stale_error}
\end{equation}
Then the gate-$k$ update
\[
\bm{\theta}^{[k]}
=
\bm{\theta}^{[k-1]}
-
\eta_{\mathrm{on}}\widehat{\Delta}_{n,k}\bm{e}_k
\]
satisfies the one-step estimate
\begin{equation}
\begin{aligned}
&L_n(\bm{\theta}^{[k]}) \\
& \leq L_n(\bm{\theta}^{[k-1]})\\
&\quad -
\eta_{\mathrm{on}}
\left[
1-\rho_{n,k}
-
\frac{L_{n,k}\eta_{\mathrm{on}}}{2}
(1+\rho_{n,k})^2
\right]
|g_{n,k}^{\mathrm{seq}}|^2.
\label{eq:online_one_step_descent}
\end{aligned}
\end{equation}
In particular, define
\begin{equation}
c_{n,k}
:=
1-\rho_{n,k}
-
\frac{L_{n,k}\eta_{\mathrm{on}}}{2}
(1+\rho_{n,k})^2 .
\label{eq:online_descent_coefficient}
\end{equation}
If
\begin{equation}
0<\eta_{\mathrm{on}}
<
\frac{2(1-\rho_{n,k})}{L_{n,k}(1+\rho_{n,k})^2}
\qquad
\text{for all } k=1,\dots,M,
\label{eq:online_coordinate_stepsize_condition}
\end{equation}
then $c_{n,k}>0$ for all gates, and one full online sweep over sample $n$
satisfies
\begin{equation}
L_n(\bm{\theta}^{[M]})
\leq
L_n(\bm{\theta}^{[0]})
-
\eta_{\mathrm{on}}
\sum_{k=1}^{M}
c_{n,k}
\left|
\frac{\partial L_n}{\partial \theta_k}(\bm{\theta}^{[k-1]})
\right|^2.
\label{eq:sample_sweep_descent}
\end{equation}
Thus, under controlled stale-adjoint error, the online sweep performs a
sequence of locally descending gate corrections for the processed sample.
\end{theorem}

The preceding result explains the algorithmic role of the online phase in a precise but limited sense. It does not imply monotonic decrease of the empirical loss
\[
F(\bm{\theta})=\frac{1}{N}\sum_{n=1}^{N}L_n(\bm{\theta})
\]
after each sample update. Different samples may have competing gradients, and an update that decreases $L_n$ may increase $L_{n'}$ for $n'\neq n$. The result instead identifies the local descent mechanism available to the online phase: each sample triggers a full sequence of gate-local corrections, and the deviation of each correction from exact sequential coordinate descent is controlled by the accumulated intra-sample displacement.

Combining Theorems~\ref{thm:online_stale_adjoint_representation}
and~\ref{thm:online_samplewise_descent} also clarifies the role of the online step size. If the gate-local signals remain bounded during one sample sweep, then
\[
|e_{n,k}| = \mathcal{O}(\eta_{\mathrm{on}})
\qquad
\text{for fixed } k.
\]
Hence the stale-adjoint perturbation vanishes in the small-step limit. Away from points where the exact sequential coordinate derivative $g_{n,k}^{\mathrm{seq}}$ is itself close to zero, the relative-error condition \eqref{eq:relative_stale_error} is therefore satisfied for sufficiently small $\eta_{\mathrm{on}}$. Near coordinate-stationary points the relative condition may fail, but the absolute error bound \eqref{eq:stale_error_bound_stepsize} still shows that the mismatch is small when the intra-sample displacement is small.

The structure identified above is distinct from the other optimizers used in the numerical comparison. Rotosolve is also coordinate-sequential, but it is objective-evaluation based and solves one-dimensional trigonometric subproblems rather than reusing a backward adjoint trajectory \cite{OstaszewskiEtAl2021Rotosolve,ParrishEtAl2019JacobiDiag}. SPSA is also inexact, but its inexactness comes from random simultaneous perturbations rather than from a stale-adjoint gate sweep \cite{Spall1998SPSA,SpallEtAl2000SPSA}. Adam, RMSProp, Adagrad, momentum, QNG, and L-BFGS-B construct global update directions at the current parameter vector, possibly with preconditioning, metric information, or quasi-Newton memory, but they do not perform a sample-wise gate-local Gauss-Seidel sweep with an amortized backward trajectory. The distinctive ingredient of online Krotov is therefore not sequentiality alone, nor inexactness alone, but the combination of sample-wise processing, gate-wise updates, and stale-adjoint amortization.

The results in this section are deliberately sample-wise. They justify the online phase as a controlled inexact sequential descent mechanism for the currently processed sample, but they do not prove that online Krotov monotonically decreases the full empirical loss, nor that it dominates batch gradient descent or quasi-Newton methods. The convergence guarantees for the full empirical objective remain those of the batch phase and the post-switch hybrid tail, as stated in Theorem~\ref{thm:batch_krotov_convergence_main}.

\section{Numerical experiments}
\label{sec:numerical}

We now examine whether the hybrid Krotov schedule remains competitive once its hyperparameters are tuned against strong optimizer baselines on fixed circuit architectures. The numerical study is deliberately organized around \emph{matched model comparisons}: for each benchmark we first fix one concrete quantum model and one concrete loss, then run independent hyperparameter sweeps for each optimizer family on exactly that same task. The figures below report, for each optimizer family, the best tested hyperparameter configuration and show the mean training loss over the available random seeds together with one-standard-deviation bands. The horizontal axis is the accepted outer iteration count of the optimizer; for native L-BFGS-B this count refers to accepted SciPy iterations rather than internal line-search probes, so that all curves are compared on a common per-update basis.

Training and test sets are now disjoint at the level of unique samples, which is essential for the four-bit parity truth table where the number of distinct inputs is only sixteen.

\subsection{Optimizer baselines and simulation cost accounting}
\label{sec:baselines}

We compare hybrid Krotov against a tuned set of standard QML optimizers drawn
from the \textsc{PennyLane} optimizer suite~\cite{BergholmEtAl2018Pennylane}
and the SciPy implementation of L-BFGS-B~\cite{VirtanenEtAl2020SciPy,ByrdEtAl1995LBFGSB,ZhuEtAl1997LBFGSB}. 
The baseline set consists of Adam~\cite{KingmaBa2015Adam},
Adagrad~\cite{DuchiEtAl2011Adagrad}, RMSProp~\cite{TielemanHinton2012RMSProp},
heavy-ball momentum~\cite{Polyak1964Momentum}, Nesterov accelerated
gradient~\cite{Nesterov1983NAG}, plain gradient descent, SPSA
\cite{Spall1998SPSA,SpallEtAl2000SPSA}, Rotosolve
\cite{OstaszewskiEtAl2021Rotosolve,ParrishEtAl2019JacobiDiag}, block-diagonal
quantum natural gradient~\cite{StokesEtAl2020QNG}, and L-BFGS-B.

A distinction between optimizer and gradient-evaluation backend is essential.
Many of these baselines are hardware-compatible in principle when their losses,
gradients, or metric information are estimated from quantum circuit
measurements. In full-state simulation, however, analytic gradients can instead
be computed by reverse-mode adjoint methods
\cite{jones2020efficientcalculationgradientsclassical,LuoEtAl2020Yao}, often
at much lower cost than parameter-shift estimation. The gate-local Krotov
implementation studied in this work belongs to this simulation regime: it uses
explicit forward states, backward-propagated costates, and their inner products.
It is therefore not a hardware-native measurement protocol in its present form.
The batch Krotov direction could still be estimated on hardware because
Theorem~\ref{thm:exact_gradient_krotov_main} identifies it with the ordinary
gradient, but such a hardware-estimated batch variant would not realize the
amortized online stale-adjoint sweep studied here.

Table~\ref{tab:complexity} summarizes the per-iteration simulation cost of the
optimizer families used in our experiments. Let $d$ denote the number of
trainable quantum parameters, $N$ the number of training samples processed per
outer iteration, $L$ the circuit depth, $m$ the L-BFGS-B memory depth, and $r$
the number of line-search probes per accepted L-BFGS-B step. The table reports
the cost of the implemented state-vector simulation setting rather than mixing
simulator-side adjoint costs with hardware measurement costs.

\begin{table*}[t]
\centering
\caption{Per-iteration simulation cost of the optimizer families considered
in this work. $d$ denotes the number of trainable quantum parameters, $N$ the
number of samples processed per outer iteration, $L$ the circuit depth, $m$
the L-BFGS-B memory depth, and $r$ the number of line-search probes per
accepted L-BFGS-B step. Costs are stated for the state-vector simulation
regime used in the experiments. Auxiliary memory refers to optimizer-level
buffers beyond the parameter vector and backend state-vector storage.}
\label{tab:complexity}
\renewcommand{\arraystretch}{1.25}
\begin{tabular}{@{}lccl@{}}
\toprule
Optimizer & State-vector simulation cost & Aux.\ memory & References \\
\midrule
Gradient descent           & $\mathcal{O}(N)$        & $\mathcal{O}(d)$      & \cite{SchuldEtAl2019Gradients,jones2020efficientcalculationgradientsclassical} \\
Momentum                   & $\mathcal{O}(N)$        & $\mathcal{O}(d)$      & \cite{Polyak1964Momentum} \\
Nesterov                   & $\mathcal{O}(N)$        & $\mathcal{O}(d)$      & \cite{Nesterov1983NAG} \\
Adagrad                    & $\mathcal{O}(N)$        & $\mathcal{O}(d)$      & \cite{DuchiEtAl2011Adagrad} \\
RMSProp                    & $\mathcal{O}(N)$        & $\mathcal{O}(d)$      & \cite{TielemanHinton2012RMSProp} \\
Adam                       & $\mathcal{O}(N)$        & $\mathcal{O}(d)$      & \cite{KingmaBa2015Adam} \\
SPSA                       & $\mathcal{O}(N)$        & $\mathcal{O}(d)$      & \cite{Spall1998SPSA,SpallEtAl2000SPSA} \\
Rotosolve                  & $\mathcal{O}(Nd)$       & $\mathcal{O}(d)$      & \cite{OstaszewskiEtAl2021Rotosolve,ParrishEtAl2019JacobiDiag} \\
L-BFGS-B                   & $\mathcal{O}(rN)$       & $\mathcal{O}(md)$     & \cite{ByrdEtAl1995LBFGSB,ZhuEtAl1997LBFGSB,LiuNocedal1989LBFGS} \\
QNG (block-diagonal)       & $\mathcal{O}(N+\sum_\ell b_\ell^2)$ & $\mathcal{O}(\sum_\ell b_\ell^2)$ & \cite{StokesEtAl2020QNG,MeyerEtAl2021FisherQML} \\
QNG (full metric)          & $\mathcal{O}(N+d^2)$    & $\mathcal{O}(d^2)$    & \cite{StokesEtAl2020QNG} \\
Hybrid Krotov (online + batch) & $\mathcal{O}(N)$    & $\mathcal{O}(d+L)$    & this work, \cite{jones2020efficientcalculationgradientsclassical}  \\
\bottomrule
\end{tabular}
\end{table*}

Table~\ref{tab:complexity} should be read as a simulation-side cost
comparison. In this regime, gradient-based first-order methods and hybrid
Krotov can all exploit adjoint information and require $\mathcal{O}(N)$
sample-wise simulations per outer iteration. Their difference is therefore not
the asymptotic number of state-vector sweeps, but how the available derivative
information is transformed into parameter updates.

The online phase of hybrid Krotov amortizes one sample-wise backward trajectory
over a full gate-wise sweep of local updates. This is the mechanism behind its
rapid early loss reduction in the experiments below. Rotosolve, by contrast,
updates coordinates through repeated objective evaluations and therefore scales
as $\mathcal{O}(Nd)$ in this simulation accounting. L-BFGS-B introduces the
line-search factor $r$, while QNG adds the cost of forming metric blocks or the
full quantum Fisher information matrix.

\subsection{Leakage-safe parity benchmark}
\label{sec:parity}

Our first benchmark is the four-bit parity classifier introduced by Farhi and Neven~\cite{FarhiNeven2018Classification} and later adopted as a tutorial example in the \textsc{PennyLane} variational-classifier demo~\cite{BergholmEtAl2018Pennylane}. We implement it both in our native optimizer pipeline and in a matched \textsc{PennyLane} reference. The quantum model uses four qubits, basis-state encoding of the four-bit input, two repeated variational layers, and measurement of $\langle Z_0\rangle$ followed by a trainable scalar bias. Each layer applies a three-parameter single-qubit Euler rotation on every qubit followed by the CNOT ring $0\!\to\!1\!\to\!2\!\to\!3\!\to\!0$. Labels are $y\in\{-1,+1\}$ with odd parity mapped to $+1$, and the training objective is the mean squared error used in the original formulation. To prevent trivial leakage through duplicated truth-table entries, we split the sixteen possible inputs into ten unique training bit strings and six unique test bit strings, which eliminates the overlap that would otherwise inflate test accuracy in short demonstrations.

For this model we tuned hybrid Krotov, native L-BFGS-B, and the \textsc{PennyLane} optimizers Adam, Adagrad, gradient descent, momentum, Nesterov momentum, RMSProp, SPSA, block-diagonal QNG, and Rotosolve. Every hyperparameter configuration was run with three independent seeds. The plotted parity curves therefore compare the best configuration \emph{within each optimizer family} rather than default settings, which is essential for a fair cross-method assessment~\cite{SchmidtEtAl2021DescendingCrowded}.

\begin{figure*}[t]
    \centering
    \includegraphics[width=\textwidth]{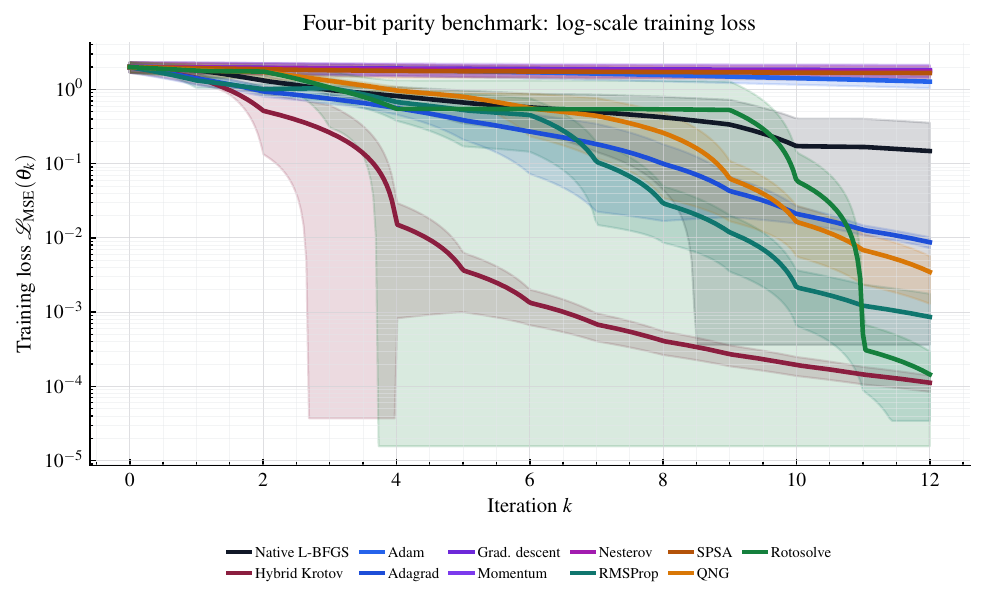}
    \caption{Leakage-safe four-bit parity benchmark~\cite{FarhiNeven2018Classification}. Shown are the best tested hyperparameter settings of each optimizer family. Curves denote the mean training loss over three seeds; shaded bands indicate one standard deviation. The native L-BFGS-B curve is recorded in accepted outer iterations, so it is directly comparable to the first-order and Krotov schedules.}
    \label{fig:parity-loss-iteration}
\end{figure*}

Figure~\ref{fig:parity-loss-iteration} exhibits three distinct regimes. First, hybrid Krotov, RMSProp, block-diagonal QNG, and Rotosolve all solve the leakage-safe task at their best settings, with hybrid Krotov and Rotosolve reaching the smallest final losses within the tested budget. Second, the tuned first-order baselines separate sharply: RMSProp and Adagrad are clearly the strongest practical \textsc{PennyLane} baselines, whereas Adam, plain gradient descent, momentum, Nesterov, and SPSA remain substantially weaker on this benchmark despite extensive tuning. Third, native L-BFGS-B is \emph{not} a dominant method on this discrete parity task; after correcting the trace accounting to compare only accepted optimizer iterations, its best tested configuration remains below the best Krotov and \textsc{PennyLane} contenders in both final loss and test accuracy.

The parity experiment functions here as a controlled calibration problem. Because the model is small and the loss is exactly matched between the native and \textsc{PennyLane} implementations, it isolates optimizer behavior from architecture confounders. In this setting the hybrid Krotov schedule is the best-performing method in this benchmark already after a modest number of iterations, while the strongest \textsc{PennyLane} baselines are RMSProp, QNG, and Rotosolve rather than Adam.

\subsection{Pérez-Salinas data re-uploading benchmark}
\label{sec:perez}
Our second benchmark is substantially less toy-like. We adopt the data re-uploading classifier introduced by Pérez-Salinas~\emph{et al.}~\cite{PerezSalinasEtAl2020DataReupload} on the two-dimensional \texttt{crown} classification problem proposed in the same work. The circuit is the entangling four-qubit, eight-layer re-uploading model with the affine classical readout head enabled. Inputs are continuous points in $[-1,1]^2$, embedded repeatedly across the circuit, and the loss is the weighted-fidelity objective used in the original formulation~\cite{PerezSalinasEtAl2020DataReupload} and mirrored exactly in our \textsc{PennyLane} reference model. For this experiment we generated $600$ samples per split, yielding $420$ training and $180$ test points under a $70/30$ split. Since the data are sampled continuously, train/test duplication is not structurally expected; we nevertheless enforced a disjoint split in the benchmark scripts for consistency with the parity protocol.

The optimizer comparison was again based on tuned families rather than defaults. We swept hybrid Krotov and the \textsc{PennyLane} optimizers Adam, Adagrad, gradient descent, momentum, Nesterov momentum, RMSProp, and SPSA with three seeds per hyperparameter combination. We also include native L-BFGS-B, truncated to the same $20$-iteration budget used in the merged comparison. A preliminary QNG study was performed separately, but the full-metric variant requires an auxiliary wire on this model~\cite{StokesEtAl2020QNG}, and the block-diagonal variant was already markedly more expensive per iteration without being competitive in accuracy; it was therefore excluded from the final large sweep.

\begin{figure*}[t]
    \centering
    \includegraphics[width=\textwidth]{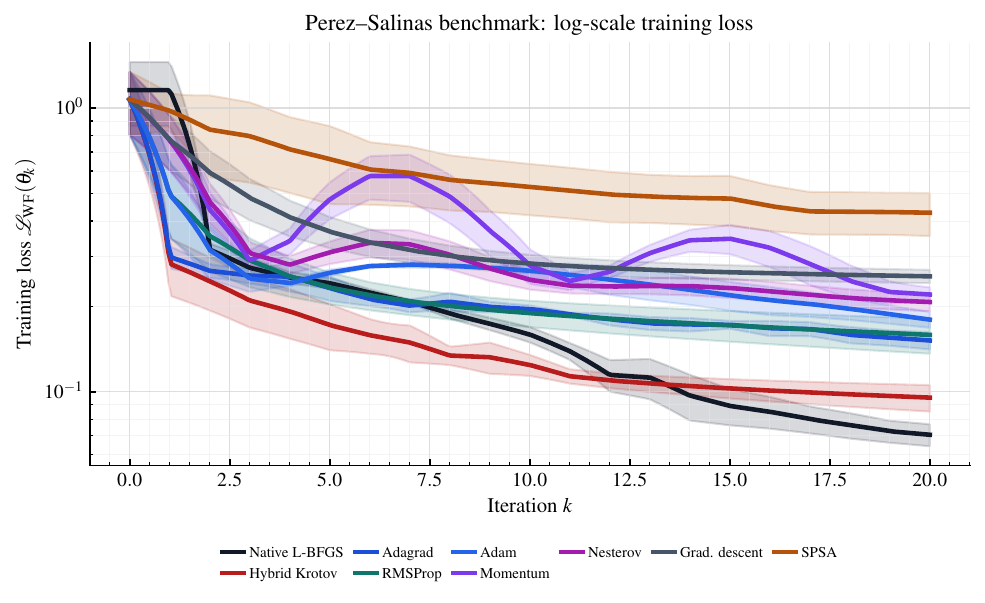}
    \caption{Pérez-Salinas data re-uploading benchmark~\cite{PerezSalinasEtAl2020DataReupload} on the \texttt{crown} problem. Each curve shows the best tested hyperparameter configuration of one optimizer family. Means and standard deviations are computed over the available seeds for that configuration; native L-BFGS-B is shown at the same $20$-iteration budget used in the sweep comparison.}
    \label{fig:perez-loss-iteration}
\end{figure*}

The behavior in Fig.~\ref{fig:perez-loss-iteration} is qualitatively different from the parity task and more informative for the intended QML application regime. Native L-BFGS-B is the strongest method at the common $20$-step budget, reaching the lowest loss and the highest test accuracy in the merged comparison; this is consistent with the well-documented strength of quasi-Newton methods on smooth moderate-dimensional problems with well-conditioned curvature~\cite{LiuNocedal1989LBFGS,ByrdEtAl1995LBFGSB}. Hybrid Krotov is nevertheless the best-performing method among the newly tuned three-seed sweep families and remains markedly stronger than the tuned \textsc{PennyLane} first-order baselines. Among those baselines, Adagrad and RMSProp emerge as the best choices, with Adam close behind but consistently weaker at its best tested setting. Momentum, Nesterov, plain gradient descent, and SPSA are clearly dominated.

Quantitatively, native L-BFGS-B reaches mean test accuracy $0.950$ and mean final loss $0.0706$ at the $20$-iteration cutoff. The best hybrid Krotov setting reaches mean test accuracy $0.935$ and mean final loss $0.0954$. The best \textsc{PennyLane} baselines remain well below this level: Adagrad attains mean test accuracy $0.832$ with mean final loss $0.1517$, RMSProp attains $0.820$ and $0.1587$, and Adam attains $0.819$ and $0.1797$. Thus, on this more structured data re-uploading benchmark, hybrid Krotov does not beat the strongest quasi-Newton baseline at equal accepted-iteration budget, but it substantially outperforms every tuned \textsc{PennyLane} first-order method in the same experiment design.

\subsection{Accuracy-cost trade-off}
\label{sec:tradeoff}

A fair assessment of the Pérez-Salinas comparison must account for the iteration-level cost differences documented in Table~\ref{tab:complexity}. Although L-BFGS-B reaches a lower loss at twenty accepted outer iterations, each such iteration involves $r$ internal line-search probes~\cite{ByrdEtAl1995LBFGSB,LiuNocedal1989LBFGS}; in our runs the average number of accepted-iteration gradient evaluations exceeded that of the first-order methods by a factor of roughly three to five. Hybrid Krotov therefore closes most of the gap to L-BFGS-B at a fraction of its per-iteration circuit-evaluation cost and at identical asymptotic memory scaling, which is the regime most relevant for scaling to larger QML models where full-batch quasi-Newton bookkeeping becomes increasingly expensive~\cite{McCleanEtAl2016Theory,CerezoEtAl2021VQA}. Conversely, the first-order \textsc{PennyLane} optimizers share hybrid Krotov's per-iteration cost but achieve noticeably worse loss and accuracy at convergence. In this sense, hybrid Krotov occupies a favourable position in the accuracy-cost plane: it matches the efficiency of first-order methods while approaching the accuracy of a tuned quasi-Newton method.

As an implementation-level check, we also recorded end-to-end wall-clock times
for the selected best configurations. These timings should not be interpreted
as backend-independent algorithmic complexity, since they depend on the native
and \textsc{PennyLane} and SciPy implementations used in the benchmark. They nevertheless
confirm that the observed performance of hybrid Krotov is not obtained at a
large runtime penalty. On the parity benchmark, hybrid Krotov was the fastest
method among the selected configurations while also reaching perfect test
accuracy. On the Pérez-Salinas benchmark, native L-BFGS-B was both fastest and
most accurate at the $20$-iteration cutoff, but hybrid Krotov was second
fastest, only about $15\%$ slower in wall-clock time, and substantially faster
than the tuned \textsc{PennyLane} first-order baselines in the same benchmark
implementation. Full wall-clock summaries are reported in Appendix~\ref{app:wallclock}.

\section{Discussion}

We have introduced the hybrid Krotov gradient descent method as a new optimizer for supervised quantum machine learning models. The benchmarks show that with its strong loss decrease in the online phase it outperforms the tested \textsc{PennyLane} first-order optimizers and stochastic baselines — Adam, Adagrad, RMSProp, momentum, Nesterov, plain gradient descent, and SPSA — across both the leakage-safe four-bit parity classifier and the Pérez-Salinas data re-uploading task on the \texttt{crown} problem. On the more structured Pérez-Salinas benchmark, hybrid Krotov closes most of the accuracy gap to the strongest tested baseline, native L-BFGS-B, while retaining the per-iteration state-vector simulation cost of a first-order method, which is the regime expected to scale most favorably to larger QML models. Theoretically, the batch phase reduces to exact full-batch gradient descent on the gate-supported parameters with standard descent and KL-convergence guarantees, and the online sweep admits a sample-wise inexact-sequential-descent characterization controlled by the stale-adjoint perturbation. Two natural follow-ups extend the method beyond the present setting. First, since the gate-local discrete adjoint construction is agnostic to the underlying state representation, scaling hybrid Krotov to larger QML models is well-suited to matrix-product-state backends such as TEBD or the recently introduced local circuit TDVP scheme~\cite{Vidal_2003,sander2025quantumcircuitsimulationlocal}. Second, the same construction transfers naturally to noisy and dissipative circuits through quantum-trajectory unravelings of Lindblad dynamics; a circuit-level adaptation of the tensor jump method~\cite{sander2025largescalestochasticsimulationopen} would extend hybrid Krotov to large-scale QML tasks under presence of noise at scalable cost.

\section{Code availability}

A full implementation of the methods including scripts to run the exact experiments of this work can be found in \cite{frohlich2026krotovoptimization}.

\section*{Acknowledgments}
This work was supported by the Einstein Research Unit on Quantum Devices. Martin Eigel acknowledges partial financial support from the German Federal Ministry of Education and Research (BMBF) under grant agreement No.~13N17160 (Q-ROM --- Quantum Read-Once-Memory: Verwandlung von klassischen Daten zu Quantenzuständen).

\bibliography{bib}

\onecolumngrid

\appendix

\section{Proof of Theorem~\ref{thm:exact_gradient_krotov_main}}

\begin{proof}
The proof adapts the reverse-mode gradient derivation of
\cite{jones2020efficientcalculationgradientsclassical} from expectation-value
objectives to sample-wise terminal-state losses and their empirical average.
Fix a sample $n$ and a gate index $k$. By definition of the forward states,
\[
|\psi_{n,M}(\bm{\theta})\rangle
=
U_M(\theta_M,x_n)\cdots U_{k+1}(\theta_{k+1},x_n)
U_k(\theta_k,x_n)
|\psi_{n,k-1}(\bm{\theta})\rangle .
\]
Differentiating with respect to $\theta_k$ yields

\[
\frac{\partial}{\partial \theta_k}
|\psi_{n,M}(\bm{\theta})\rangle
=
U_M(\theta_M,x_n)\cdots U_{k+1}(\theta_{k+1},x_n)
\frac{\partial U_k}{\partial \theta_k}(\theta_k,x_n)
|\psi_{n,k-1}(\bm{\theta})\rangle .
\]

Since $L_n(\bm{\theta})$ depends on the circuit only through the final state
and the terminal costate is defined by
\[
\delta L_n(\bm{\theta})
=
2\,\mathrm{Re}
\langle
\chi_{n,M}(\bm{\theta})
\mid
\delta\psi_{n,M}(\bm{\theta})
\rangle ,
\]
we may apply this identity to the variation induced by changing $\theta_k$.
Thus
\[
\frac{\partial L_n}{\partial \theta_k}(\bm{\theta})
=
2\,\mathrm{Re}
\left\langle
\chi_{n,M}(\bm{\theta})
\middle|
\frac{\partial \psi_{n,M}}{\partial \theta_k}(\bm{\theta})
\right\rangle .
\]
Substituting the previous expression gives
\[
\frac{\partial L_n}{\partial \theta_k}(\bm{\theta})
=
2\,\mathrm{Re}
\left\langle
\chi_{n,M}(\bm{\theta})
\middle|
U_M(\theta_M,x_n)\cdots U_{k+1}(\theta_{k+1},x_n)
\frac{\partial U_k}{\partial \theta_k}(\theta_k,x_n)
\middle|
\psi_{n,k-1}(\bm{\theta})
\right\rangle .
\]

Using the definition of the backward-propagated costate,
\[
|\chi_{n,k}^{\mathrm{after}}(\bm{\theta})\rangle
=
U_{k+1}(\theta_{k+1},x_n)^\dagger\cdots
U_M(\theta_M,x_n)^\dagger
|\chi_{n,M}(\bm{\theta})\rangle ,
\]
we obtain
\[
\langle \chi_{n,M}(\bm{\theta})|
U_M(\theta_M,x_n)\cdots U_{k+1}(\theta_{k+1},x_n)
=
\langle \chi_{n,k}^{\mathrm{after}}(\bm{\theta})| .
\]
Hence
\[
\frac{\partial L_n}{\partial \theta_k}(\bm{\theta})
=
2\,\mathrm{Re}
\left\langle
\chi_{n,k}^{\mathrm{after}}(\bm{\theta})
\middle|
\frac{\partial U_k}{\partial \theta_k}(\theta_k,x_n)
\middle|
\psi_{n,k-1}(\bm{\theta})
\right\rangle .
\]

Now use the fixed-generator parametrization
\[
U_k(\theta_k,x_n)
=
\exp\!\left(
-\frac{i}{2}\alpha_k(\theta_k,x_n)G_k
\right).
\]
Since $G_k$ is independent of $\theta_k$, the parameter dependence enters only
through the scalar angle map $\alpha_k(\theta_k,x_n)$. Therefore
\[
\frac{\partial U_k}{\partial \theta_k}(\theta_k,x_n)
=
\frac{\partial \alpha_k}{\partial \theta_k}(\theta_k,x_n)
\left(-\frac{i}{2}G_k\right)
U_k(\theta_k,x_n).
\]
Substituting this identity yields
\begin{align*}
\frac{\partial L_n}{\partial \theta_k}(\bm{\theta})
&=
\frac{\partial \alpha_k}{\partial \theta_k}(\theta_k,x_n)\,
2\,\mathrm{Re}
\left\langle
\chi_{n,k}^{\mathrm{after}}(\bm{\theta})
\middle|
\left(-\frac{i}{2}G_k\right)
U_k(\theta_k,x_n)
\middle|
\psi_{n,k-1}(\bm{\theta})
\right\rangle \\
&=
\frac{\partial \alpha_k}{\partial \theta_k}(\theta_k,x_n)\,
2\,\mathrm{Re}
\left\langle
\chi_{n,k}^{\mathrm{after}}(\bm{\theta})
\middle|
\left(-\frac{i}{2}G_k\right)
\middle|
\psi_{n,k}(\bm{\theta})
\right\rangle ,
\end{align*}
because
\[
|\psi_{n,k}(\bm{\theta})\rangle
=
U_k(\theta_k,x_n)|\psi_{n,k-1}(\bm{\theta})\rangle .
\]
This proves the sample-wise identity
\[
\Delta_{n,k}(\bm{\theta})
=
\frac{\partial L_n}{\partial \theta_k}(\bm{\theta}) .
\]

Averaging over $n=1,\dots,N$ gives
\[
\bar{\Delta}_k(\bm{\theta})
=
\frac{1}{N}\sum_{n=1}^N \Delta_{n,k}(\bm{\theta})
=
\frac{1}{N}\sum_{n=1}^N
\frac{\partial L_n}{\partial \theta_k}(\bm{\theta})
=
\frac{\partial F}{\partial \theta_k}(\bm{\theta}) .
\]
Since this holds for every $k=1,\dots,M$, the batch Krotov direction coincides
with the gradient of the empirical loss with respect to the gate-supported
parameters.
\end{proof}

\section{Proof of Theorem~\ref{thm:batch_krotov_convergence_main}}

\begin{proof}
By assumption, the batch Krotov direction equals the exact gradient of $F$.
Hence the batch Krotov iteration is exactly gradient descent,
\[
\bm{\theta}^{(m+1)}
=
\bm{\theta}^{(m)}
-
\eta\nabla F(\bm{\theta}^{(m)}).
\]
Since $F$ has an $L$-Lipschitz continuous gradient on the sublevel set
$\mathcal S_0$, the descent lemma gives
\[
F(\bm{\theta}^{(m+1)})
\leq
F(\bm{\theta}^{(m)})
+
\left\langle
\nabla F(\bm{\theta}^{(m)}),
\bm{\theta}^{(m+1)}-\bm{\theta}^{(m)}
\right\rangle
+
\frac{L}{2}
\|\bm{\theta}^{(m+1)}-\bm{\theta}^{(m)}\|^2 .
\]
Substituting
\[
\bm{\theta}^{(m+1)}-\bm{\theta}^{(m)}
=
-\eta\nabla F(\bm{\theta}^{(m)})
\]
yields
\begin{align*}
F(\bm{\theta}^{(m+1)})
&\leq
F(\bm{\theta}^{(m)})
-
\eta\|\nabla F(\bm{\theta}^{(m)})\|^2
+
\frac{L\eta^2}{2}
\|\nabla F(\bm{\theta}^{(m)})\|^2 \\
&=
F(\bm{\theta}^{(m)})
-
\eta
\left(
1-\frac{L\eta}{2}
\right)
\|\nabla F(\bm{\theta}^{(m)})\|^2 .
\end{align*}
Because $0<\eta<2/L$, the coefficient
\[
c
:=
\eta
\left(
1-\frac{L\eta}{2}
\right)
\]
is strictly positive. This proves the descent estimate and, in particular,
the monotonic nonincrease of $F(\bm{\theta}^{(m)})$.

Summing the descent estimate from $m=0$ to $M$ gives
\[
F(\bm{\theta}^{(M+1)})
\leq
F(\bm{\theta}^{(0)})
-
c\sum_{m=0}^{M}
\|\nabla F(\bm{\theta}^{(m)})\|^2 .
\]
Since $F$ is bounded from below, say by $F_{\inf} \in \mathbb{R}$, it follows that
\[
c\sum_{m=0}^{M}
\|\nabla F(\bm{\theta}^{(m)})\|^2
\leq
F(\bm{\theta}^{(0)})-F_{\inf}.
\]
Letting $M\to\infty$ yields
\[
\sum_{m=0}^{\infty}
\|\nabla F(\bm{\theta}^{(m)})\|^2
<\infty .
\]
Consequently,
\[
\|\nabla F(\bm{\theta}^{(m)})\|\to 0 .
\]

Now let $\bm{\theta}^{\star}$ be an accumulation point of the sequence. Then
there exists a subsequence $\bm{\theta}^{(m_j)}$ such that
\[
\bm{\theta}^{(m_j)}
\to
\bm{\theta}^{\star}.
\]
Since $\nabla F$ is continuous on $\mathcal S_0$,
\[
\nabla F(\bm{\theta}^{\star})
=
\lim_{j\to\infty}
\nabla F(\bm{\theta}^{(m_j)})
=
0 .
\]
Thus every accumulation point is stationary.

If, in addition, the iterates are bounded and $F$ satisfies the
Kurdyka-\L{}ojasiewicz property, the standard KL convergence theorem for
descent methods applies, and the whole sequence converges to a single
stationary point.

Finally, the same argument applies to the post-switch phase of a hybrid run.
If the method switches to the batch update after a finite time
$T_{\mathrm{sw}}$, define the shifted sequence
\[
\widetilde{\bm{\theta}}^{(j)}
:=
\bm{\theta}^{(T_{\mathrm{sw}}+j)},
\qquad
j\geq 0.
\]
After the switch, this shifted sequence satisfies
\[
\widetilde{\bm{\theta}}^{(j+1)}
=
\widetilde{\bm{\theta}}^{(j)}
-
\eta\nabla F(\widetilde{\bm{\theta}}^{(j)}),
\]
and is therefore exactly a batch Krotov sequence initialized at
$\bm{\theta}^{(T_{\mathrm{sw}})}$. Applying the preceding argument to this
shifted sequence proves the stated hybrid-tail conclusion.
\end{proof}

\section{Proof of Theorem~\ref{thm:online_stale_adjoint_representation}}

\begin{proof}
Fix a sample $n$ and a gate index $k$. By definition of the mixed adjoint
signal,
\[
\widehat{\Delta}_{n,k}
=
\mathcal{D}_{n,k}(\bm{\theta}^{[0]},\bm{\theta}^{[k-1]}).
\]
The exact sequential coordinate-gradient signal at the partially updated
parameter vector is
\[
g_{n,k}^{\mathrm{seq}}
=
\frac{\partial L_n}{\partial \theta_k}(\bm{\theta}^{[k-1]}).
\]
By the diagonal identity \eqref{eq:mixed_signal_diagonal_identity}, this can be
written as
\[
g_{n,k}^{\mathrm{seq}}
=
\mathcal{D}_{n,k}(\bm{\theta}^{[k-1]},\bm{\theta}^{[k-1]}).
\]
Therefore
\[
\widehat{\Delta}_{n,k}
-
\frac{\partial L_n}{\partial \theta_k}(\bm{\theta}^{[k-1]})
=
\mathcal{D}_{n,k}(\bm{\theta}^{[0]},\bm{\theta}^{[k-1]})
-
\mathcal{D}_{n,k}(\bm{\theta}^{[k-1]},\bm{\theta}^{[k-1]}).
\]
Define
\[
e_{n,k}
:=
\widehat{\Delta}_{n,k}
-
\frac{\partial L_n}{\partial \theta_k}(\bm{\theta}^{[k-1]}).
\]
Applying the assumed Lipschitz bound with
\[
\bm a=\bm{\theta}^{[0]},
\qquad
\bm b=\bm{\theta}^{[k-1]},
\]
gives
\[
|e_{n,k}|
\leq
C_{n,k}
\|\bm{\theta}^{[0]}-\bm{\theta}^{[k-1]}\|
=
C_{n,k}
\|\bm{\theta}^{[k-1]}-\bm{\theta}^{[0]}\|.
\]
This proves \eqref{eq:stale_error_bound_displacement}.

It remains to express this displacement in terms of the previous intra-sample
updates. From the recursive definition of the online update,
\[
\bm{\theta}^{[j]}
=
\bm{\theta}^{[j-1]}
-
\eta_{\mathrm{on}}
\widehat{\Delta}_{n,j}\bm e_j,
\qquad
j=1,\dots,k-1,
\]
we obtain
\[
\bm{\theta}^{[k-1]}-\bm{\theta}^{[0]}
=
-\eta_{\mathrm{on}}
\sum_{j=1}^{k-1}
\widehat{\Delta}_{n,j}\bm e_j .
\]
Since the vectors $\bm e_j$ are orthonormal in parameter space,
\[
\|\bm{\theta}^{[k-1]}-\bm{\theta}^{[0]}\|
=
\eta_{\mathrm{on}}
\left(
\sum_{j=1}^{k-1}
|\widehat{\Delta}_{n,j}|^2
\right)^{1/2}.
\]
Substituting this expression into
\eqref{eq:stale_error_bound_displacement} gives
\[
|e_{n,k}|
\leq
C_{n,k}\eta_{\mathrm{on}}
\left(
\sum_{j=1}^{k-1}
|\widehat{\Delta}_{n,j}|^2
\right)^{1/2}.
\]
This proves \eqref{eq:stale_error_bound_stepsize}.
\end{proof}

\section{Proof of Theorem~\ref{thm:online_samplewise_descent}}

\begin{proof}
Fix a sample $n$ and a gate index $k$. Set
\[
g
:=
g_{n,k}^{\mathrm{seq}}
=
\frac{\partial L_n}{\partial \theta_k}(\bm{\theta}^{[k-1]}),
\qquad
e:=e_{n,k}.
\]
Then
\[
\widehat{\Delta}_{n,k}=g+e.
\]
The online update changes only the $k$-th coordinate:
\[
\bm{\theta}^{[k]}
=
\bm{\theta}^{[k-1]}
-
\eta_{\mathrm{on}}(g+e)\bm e_k .
\]
Thus the scalar coordinate displacement is
\[
s
=
-\eta_{\mathrm{on}}(g+e).
\]

By the coordinate-wise smoothness assumption in the theorem statement,
\[
L_n(\bm{\theta}^{[k]})
\leq
L_n(\bm{\theta}^{[k-1]})
+
s\,g
+
\frac{L_{n,k}}{2}s^2 .
\]
Substituting $s=-\eta_{\mathrm{on}}(g+e)$ gives
\[
L_n(\bm{\theta}^{[k]})
\leq
L_n(\bm{\theta}^{[k-1]})
-
\eta_{\mathrm{on}}g(g+e)
+
\frac{L_{n,k}\eta_{\mathrm{on}}^2}{2}(g+e)^2 .
\]
Using the relative stale-adjoint error bound
\[
|e|\leq \rho_{n,k}|g|,
\qquad
0\leq \rho_{n,k}<1,
\]
we estimate the linear term by
\[
-g(g+e)
=
-g^2-ge
\leq
-g^2+|g||e|
\leq
-(1-\rho_{n,k})g^2.
\]
Moreover,
\[
|g+e|
\leq
|g|+|e|
\leq
(1+\rho_{n,k})|g|,
\]
and therefore
\[
(g+e)^2
\leq
(1+\rho_{n,k})^2g^2.
\]
Combining these estimates yields
\begin{align*}
L_n(\bm{\theta}^{[k]})
&\leq
L_n(\bm{\theta}^{[k-1]})
-
\eta_{\mathrm{on}}(1-\rho_{n,k})g^2
+
\frac{L_{n,k}\eta_{\mathrm{on}}^2}{2}
(1+\rho_{n,k})^2g^2 \\
&=
L_n(\bm{\theta}^{[k-1]})
-
\eta_{\mathrm{on}}
\left[
1-\rho_{n,k}
-
\frac{L_{n,k}\eta_{\mathrm{on}}}{2}
(1+\rho_{n,k})^2
\right]
g^2 .
\end{align*}
Restoring
\[
g=g_{n,k}^{\mathrm{seq}}
\]
gives the one-step estimate
\[
L_n(\bm{\theta}^{[k]})
\leq
L_n(\bm{\theta}^{[k-1]})
-
\eta_{\mathrm{on}}
c_{n,k}
|g_{n,k}^{\mathrm{seq}}|^2,
\]
where
\[
c_{n,k}
=
1-\rho_{n,k}
-
\frac{L_{n,k}\eta_{\mathrm{on}}}{2}
(1+\rho_{n,k})^2 .
\]
Equivalently, this is \eqref{eq:online_one_step_descent}. The step-size
condition in the theorem statement is precisely the condition $c_{n,k}>0$.

Applying the one-step estimate for $k=1,\dots,M$ and summing gives
\[
\sum_{k=1}^{M}
\left[
L_n(\bm{\theta}^{[k]})
-
L_n(\bm{\theta}^{[k-1]})
\right]
\leq
-
\eta_{\mathrm{on}}
\sum_{k=1}^{M}
c_{n,k}
|g_{n,k}^{\mathrm{seq}}|^2.
\]
The left-hand side telescopes:
\[
\sum_{k=1}^{M}
\left[
L_n(\bm{\theta}^{[k]})
-
L_n(\bm{\theta}^{[k-1]})
\right]
=
L_n(\bm{\theta}^{[M]})
-
L_n(\bm{\theta}^{[0]}).
\]
Hence
\[
L_n(\bm{\theta}^{[M]})
\leq
L_n(\bm{\theta}^{[0]})
-
\eta_{\mathrm{on}}
\sum_{k=1}^{M}
c_{n,k}
|g_{n,k}^{\mathrm{seq}}|^2.
\]
Using
\[
g_{n,k}^{\mathrm{seq}}
=
\frac{\partial L_n}{\partial \theta_k}
(\bm{\theta}^{[k-1]})
\]
gives exactly \eqref{eq:sample_sweep_descent}. Since the step-size condition
implies $c_{n,k}>0$ for all $k$, the right-hand side is a descent estimate for
the processed sample, unless all relevant sequential coordinate derivatives
vanish.
\end{proof}

\section{Wall-clock timing summaries}
\label{app:wallclock}

This appendix reports end-to-end wall-clock times for the selected best
hyperparameter configurations in the two benchmark sweeps. The timings are
implementation-dependent and should not be read as backend-independent
algorithmic complexity. They were recorded to verify that the loss and accuracy
improvements reported in the main text were not obtained at a disproportionate
runtime cost.

\begin{table}[ht]
\centering
\caption{Wall-clock times for the selected best configurations in the
leakage-safe four-bit parity benchmark. Times are averaged over three seeds.}
\label{tab:parity_wallclock}
\renewcommand{\arraystretch}{1.15}
\begin{tabular}{@{}lccc@{}}
\toprule
Optimizer & Wall time [s] & Test accuracy & Final loss \\
\midrule
Hybrid Krotov & $1.17 \pm 0.09$ & $100.00\%$ & $0.0001$ \\
Native L-BFGS-B & $1.21 \pm 0.16$ & $83.33\%$ & $0.1478$ \\
PennyLane SPSA & $2.46 \pm 0.09$ & $50.00\%$ & $1.6631$ \\
PennyLane RMSProp & $2.62 \pm 0.20$ & $100.00\%$ & $0.0009$ \\
PennyLane Adagrad & $2.72 \pm 0.22$ & $100.00\%$ & $0.0087$ \\
PennyLane Adam & $2.73 \pm 0.10$ & $50.00\%$ & $1.2724$ \\
PennyLane Nesterov & $2.75 \pm 0.16$ & $50.00\%$ & $1.7583$ \\
PennyLane Momentum & $2.80 \pm 0.06$ & $50.00\%$ & $1.7701$ \\
PennyLane Gradient Descent & $2.81 \pm 0.18$ & $50.00\%$ & $1.8228$ \\
PennyLane Rotosolve & $39.46 \pm 1.73$ & $100.00\%$ & $0.0001$ \\
PennyLane QNG & $143.60 \pm 19.33$ & $100.00\%$ & $0.0035$ \\
\bottomrule
\end{tabular}
\end{table}

\begin{table}[ht]
\centering
\caption{Wall-clock times for the selected best configurations in the
Pérez-Salinas \texttt{crown} benchmark. Times are averaged over the available
seeds.}
\label{tab:perez_wallclock}
\renewcommand{\arraystretch}{1.15}
\begin{tabular}{@{}lcccc@{}}
\toprule
Optimizer & Wall time [s] & Test accuracy & Final loss & Seeds \\
\midrule
Native L-BFGS-B & $202.93$ & $95.00\%$ & $0.0706$ & $3$ \\
Hybrid Krotov & $232.05$ & $93.52\%$ & $0.0954$ & $3$ \\
PennyLane SPSA & $645.05$ & $51.67\%$ & $0.4277$ & $3$ \\
PennyLane Nesterov & $786.99$ & $68.15\%$ & $0.2073$ & $3$ \\
PennyLane RMSProp & $791.66$ & $82.04\%$ & $0.1587$ & $3$ \\
PennyLane Momentum & $799.13$ & $71.48\%$ & $0.2203$ & $3$ \\
PennyLane Adam & $800.22$ & $81.85\%$ & $0.1797$ & $3$ \\
PennyLane Adagrad & $805.60$ & $83.15\%$ & $0.1517$ & $3$ \\
PennyLane Gradient Descent & $810.69$ & $54.81\%$ & $0.2553$ & $3$ \\
\bottomrule
\end{tabular}
\end{table}

\end{document}